\documentclass[reprint,twocolumn,superscriptaddress,preprintnumbers]{revtex4}%
\usepackage[a4paper]{geometry}
\usepackage{graphicx}
\usepackage{bm}
\usepackage{latexsym}
\usepackage{epsf}
\usepackage{rotating}
\usepackage{epsfig,graphics,rotate,xcolor}
\usepackage{wrapfig}
\usepackage{amssymb}
\usepackage{amsmath}
\usepackage{amsfonts}
\usepackage{array,hhline,dcolumn}%
\usepackage[normalem]{ulem}
\providecommand{\U}[1]{\protect\rule{.1in}{.1in}}
\begin{document}

\title{Measurements of Inclusive Muon Neutrino and Antineutrino Charged Current Differential Cross Sections on Argon in the NuMI Antineutrino Beam}

\author{R.~Acciarri}
\affiliation{Fermi National Accelerator Laboratory, Batavia, IL 60510}

\author{C.~Adams}
\affiliation{Yale University, New Haven, CT 06520}


\author{J.~Asaadi}
\affiliation{Syracuse University, Syracuse, NY 13244}

\author{B.~Baller}
\affiliation{Fermi National Accelerator Laboratory, Batavia, IL 60510}

\author{T.~Bolton}
\affiliation{Kansas State University, Manhattan, KS 66506}

\author{C.~Bromberg}
\affiliation{Michigan State University, East Lansing, MI 48824}

\author{F.~Cavanna}
\affiliation{Yale University, New Haven, CT 06520}
\affiliation{Universita dell'Aquila e INFN, L'Aquila, Italy}

\author{E.~Church}
\affiliation{Yale University, New Haven, CT 06520}

\author{D.~Edmunds}
\affiliation{Michigan State University, East Lansing, MI 48824}

\author{A.~Ereditato}
\affiliation{University of Bern, Bern, Switzerland}

\author{S.~Farooq}
\affiliation{Kansas State University, Manhattan, KS 66506}

\author{B.~Fleming}
\affiliation{Yale University, New Haven, CT 06520}

\author{H.~Greenlee}
\affiliation{Fermi National Accelerator Laboratory, Batavia, IL 60510}


\author{R.~Hatcher}
\affiliation{Fermi National Accelerator Laboratory, Batavia, IL 60510}

\author{G.~Horton-Smith}
\affiliation{Kansas State University, Manhattan, KS 66506}

\author{C.~James}
\affiliation{Fermi National Accelerator Laboratory, Batavia, IL 60510}

\author{E.~Klein}
\affiliation{Yale University, New Haven, CT 06520}

\author{K.~Lang}
\affiliation{The University of Texas at Austin, Austin, TX 78712}

\author{P.~Laurens}
\affiliation{Michigan State University, East Lansing, MI 48824}


\author{R.~Mehdiyev}
\affiliation{The University of Texas at Austin, Austin, TX 78712}

\author{B.~Page}
\affiliation{Michigan State University, East Lansing, MI 48824}

\author{O.~Palamara}
\affiliation{Yale University, New Haven, CT 06520}
\affiliation{INFN - Laboratori Nazionali del Gran Sasso, Assergi, Italy}

\author{K.~Partyka}
\affiliation{Yale University, New Haven, CT 06520}

\author{G.~Rameika}
\affiliation{Fermi National Accelerator Laboratory, Batavia, IL 60510}

\author{B.~Rebel}
\affiliation{Fermi National Accelerator Laboratory, Batavia, IL 60510}

\author{M.~Soderberg}
\affiliation{Fermi National Accelerator Laboratory, Batavia, IL 60510}
\affiliation{Syracuse University, Syracuse, NY 13244}

\author{J.~Spitz}
\affiliation{Yale University, New Haven, CT 06520}

\author{A.M.~Szelc}
\affiliation{Yale University, New Haven, CT 06520}

\author{M.~Weber}
\affiliation{University of Bern, Bern, Switzerland}

\author{T.~Yang}
\affiliation{Fermi National Accelerator Laboratory, Batavia, IL 60510}

\author{G.P.~Zeller}
\affiliation{Fermi National Accelerator Laboratory, Batavia, IL 60510}

\collaboration{The ArgoNeuT Collaboration}
\noaffiliation

\preprint{FERMILAB-PUB-14-104-E}

\begin{abstract} 
The ArgoNeuT collaboration presents measurements of inclusive muon neutrino and antineutrino charged current differential cross sections on argon in the Fermilab NuMI beam operating in the low energy antineutrino mode. The results are reported in terms of outgoing muon angle and momentum at a mean neutrino energy of 9.6 GeV (neutrinos) and 3.6 GeV (antineutrinos), in the range $0^\circ < \theta_\mu < 36^\circ$ and $0 < p_\mu < 25$~GeV/$c$, for both neutrinos and antineutrinos.
\end{abstract}

\maketitle

The liquid argon time projection chamber (LArTPC)~\cite{rubbia} is a detection technique that provides three-dimensional imaging of charged particle interactions with fine spatial resolution, ideal for the study of neutrino interactions~\cite{wanfprd,josh,Karagiorgi:2013cwa} and searches for rare phenomena such as proton decay. In the U.S., ArgoNeuT (Argon Neutrino Test)~\cite{flavio} and MicroBooNE~\cite{Chen:2007ae} represent an important development towards the realization of multikiloton-scale LArTPC detectors. With the recent decision for the future U.S. long-baseline neutrino detector to be a large LArTPC~\cite{lbnetech}, it is crucial to understand both neutrino and antineutrino interactions on argon for future CP violation searches. Neutrino cross sections have been measured on a variety of targets~\cite{pdg,t2k2, minervanu, minervaanu}; however, measurements on liquid argon are virtually non-existent~\cite{josh}. In this Letter, we present flux-averaged $\nu_\mu$ and $\overline{\nu}_\mu$ charged current (CC) differential cross sections as measured with the ArgoNeuT LArTPC exposed to the NuMI beam~\cite{numibeam} in a neutrino kinematic range relevant for current and future accelerator-based neutrino experiments.
The integrated cross sections are also reported. This paper, representing the first measurement of $\overline{\nu}_{\mu}$-Ar cross sections, follows the analysis approach of our previous measurement~\cite{josh}, differing in its study of the NuMI beam operating in antineutrino mode in place of neutrino mode.

The NuMI beam can be configured to produce a $\nu_{\mu}$ or $\overline{\nu}_{\mu}$ enhanced beam by changing the polarity of the current in the magnetic focusing horns. 
The neutrinos in the antineutrino beam have a broad energy spectrum and originate from the decay of forward-going $\pi^+$'s and $K^{+}$'s that were not defocused by the horns. The antineutrinos from focused $\pi^-$'s come in a narrowly distributed, lower energy spectrum. 
Neutrino interactions comprise almost 60\% of all the neutrino/antineutrino-induced events in the detector in antineutrino mode~\cite{minosnue}. This fact enables the measurements of both $\nu_\mu$ and $\overline{\nu}_\mu$ CC differential cross sections using this data sample.

ArgoNeuT~\cite{flavio} is the first LArTPC to take data in a low energy neutrino beam.
 ArgoNeuT collected neutrino and antineutrino events in Fermilab's NuMI beamline at the MINOS near detector hall from September, 2009 through February, 2010. The MINOS near detector is a 1 kton magnetized steel/scintillator tracking/sampling calorimeter~\cite{minossteel}. The measurements reported in this Letter are based on data taken with the NuMI beamline operating in antineutrino mode, corresponding to 1.20$\times10^{20}$~protons on target (POT) during which both the ArgoNeuT and MINOS near detector were operational. 
ArgoNeuT instruments a set of two wire planes at the edge of a 170~liter TPC in order to detect neutrino-induced particle tracks. The TPC's active volume is 47 $\times$ 40 $\times$ 90~cm$^3$ (horizontal drift dimension $\times$ height $\times$ length). Ionization electrons created by charged particles traversing the medium are drifted toward the sensing wire planes by a 481 V/cm electric field.  


Neutrino event reconstruction uses the LArSoft~\cite{larsoft} software package. The automated reconstruction procedure finds hits, and then defines clusters as hits in the same wire plane that may be grouped together due to proximity to one another. Line-like clusters are then found in each of the two wire plane view by projecting short ``seed'' clusters formed in regions of low hit density into higher hit density regions. Hits are appended to the seed cluster if the projected hit position and hit charge are consistent with the seed cluster hits. 
Three dimensional tracks are constructed from pairs of these line-like clusters in each plane for lines that have endpoints common in time. 

Since ArgoNeuT is too small to completely contain GeV-scale muons, the magnetized MINOS near detector provides the measurements of muon momentum and the sign of the muon charge. 
This sign-selection is especially important for measurements in the NuMI antineutrino beam. The front face of MINOS is approximately 1.5~m downstream of ArgoNeuT.
After tracking final state muons in ArgoNeuT, an attempt is made to match the three dimensional tracks that leave the ArgoNeuT TPC with muons that have been reconstructed in MINOS and have a hit within 20~cm of the upstream face of the MINOS detector. This matching criterion is based on the radial and angular differences between the projected-to-MINOS ArgoNeuT track and the candidate MINOS track and a consideration of the expected muon multiple scattering. 
In the ArgoNeuT detector itself, the muon energy loss before reaching MINOS is calculated from the distance the muon traverses the liquid argon. The muon angle is measured in the ArgoNeuT TPC. If the muon is reconstructed in the MINOS near detector, the muon momentum is measured by MINOS from the track curvature or its range depending on whether the muon stops in MINOS. The muon angular resolution determined by ArgoNeuT is 1-3$^\circ$~\cite{spitzthesis}, and the momentum resolution determined by MINOS is 5-10\%~\cite{flux}.

The reconstructed neutrino interaction vertex is required to be inside of the ArgoNeuT fiducial volume, defined as a rectangular box shaped as follows: the boundary from the plane comprising the inner sense wires and that of the cathode plane is 3~cm, the boundary from the top and bottom of the TPC is 4~cm, and the distances from the upstream (downstream) end of the TPC are 6~cm (4~cm). 
Neutrino-induced through-going  muons that pass the fiducial volume requirement due to possible inefficiency in vertex reconstruction are removed through visual scanning of events. The track matching criterion for CC $\nu_\mu$ ($\overline{\nu}_\mu$) interactions along with a requirement that the reconstructed and matched MINOS track is negatively (positively) charged represent the only other selection criteria used in this analysis. After all the selection requirements, 1676 (1605) $\nu_\mu$ ($\overline\nu_\mu$) CC candidate events remain in the data sample.

Simulated neutrino events are used to estimate the detection efficiency, measurement resolution, mismatched track rate, and neutral-current background.
An example of a mismatched track is one where the simulated ArgoNeuT muon is matched in the MINOS detector but has the wrong reconstructed sign. 
The simulation of neutrino interactions in ArgoNeuT employs a {\sc geant4}-based~\cite{geant4} detector model and particle propagation software in combination with the {\sc genie} neutrino event generator~\cite{genie}. 
The propagation of particles in the MINOS near detector is simulated with {\sc geant3}~\cite{geant3}.
A standalone version of MINOS simulation and reconstruction is used to characterize the matching of tracks passing from ArgoNeuT into MINOS. In this study, we generated neutrino interactions in the LArTPC and muons in MINOS (both from ArgoNeuT and surrounding rock) to calculate the efficiency and background rates. The number of mismatched track and neutral-current background events is 97 (141) for the $\nu_{\mu}$ ($\overline{\nu}_{\mu}$) sample.


Following event selection, the differential cross section in terms of a measured variable $u$ in bin $i$ is extracted using
\begin{equation}
\frac{d\sigma (u_{i})}{du }=\frac{N_{\mathrm{measured},i}-N_{\mathrm{background},i}}{\Delta u_{i}~\epsilon_i~N_{\mathrm{targ}}~\Phi}~,
\label{diffeq}
\end{equation}
where $N_{\mathrm{measured},i}$ represents the number of events in data passing analysis selection, $N_{\mathrm{background},i}$ is the number of expected background events, $\Delta u_{i}$ is the bin width, $\epsilon_i$ is the detection efficiency, $N_{\mathrm{targ}}$ is the number of argon nuclei in the fiducial volume, and $\Phi$ is the total neutrino flux exposure. 
The variable $u$ represents the outgoing muon angle with respect to the initial neutrino direction ($\theta_\mu$) or the momentum ($p_\mu$). 

The $\theta_\mu$ and $p_\mu$ distributions are corrected for efficiency on a bin-by-bin basis after subtracting the estimated background contributions according to Eq.~\ref{diffeq}. A $\nu_\mu$ ($\overline{\nu}_\mu$) CC event that originates in the ArgoNeuT fiducial volume enters the signal sample after ArgoNeuT-MINOS reconstruction, track matching, and selection 42.0\% (59.0\%) of the time on average.
This is due to muon acceptance between ArgoNeuT and MINOS, track matching between ArgoNeuT and MINOS, vertex reconstruction inefficiencies in ArgoNeuT, track reconstruction inefficiencies in both detectors, event selection efficiency, and the effect of detector resolution. In general, neutrino interactions have higher energy hadronic activity than antineutrino interactions, resulting in a reduced efficiency for neutrino events and a larger systematic error on the fiducial volume.  

\begin{figure}[htp]
\includegraphics[width=1.0\linewidth]{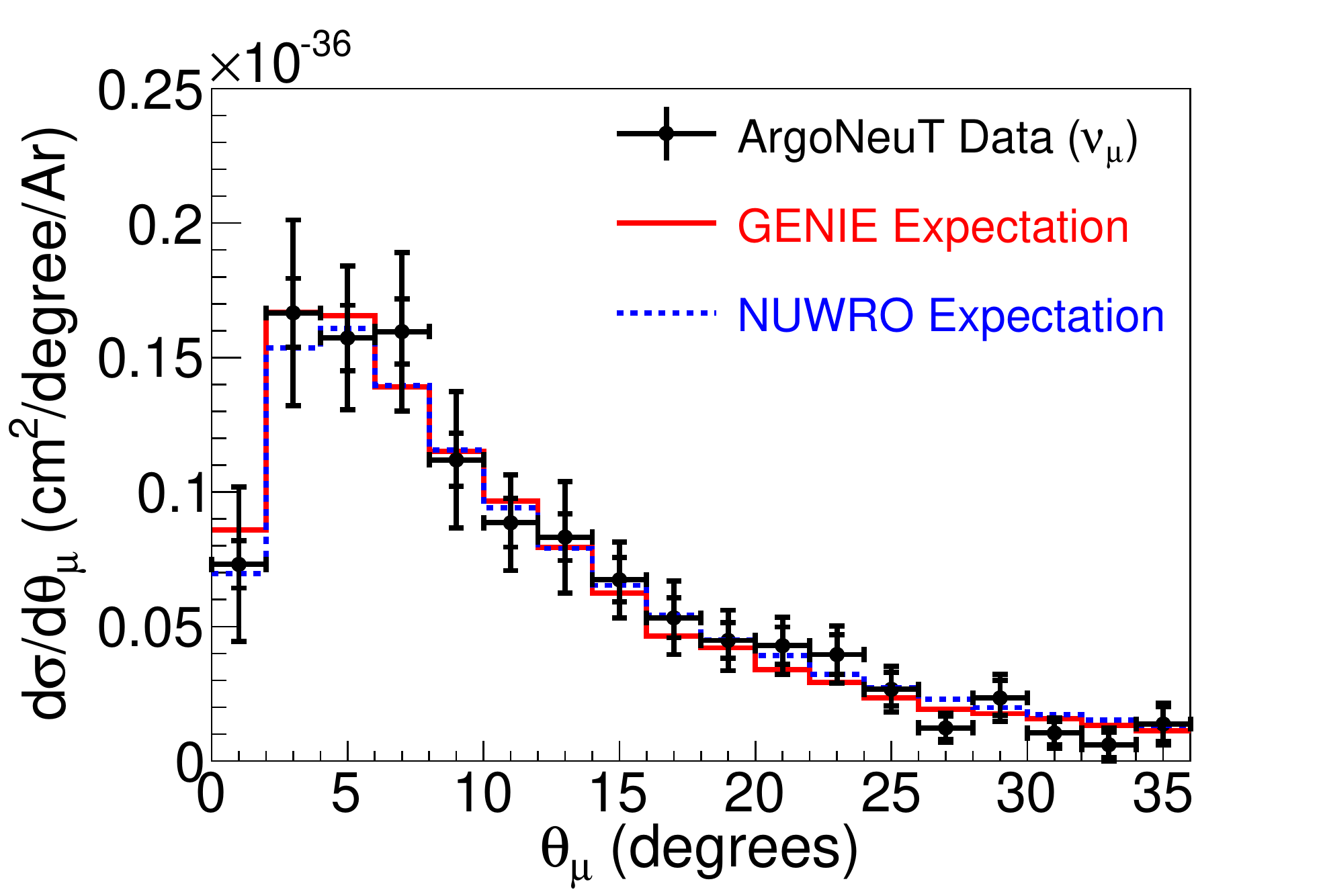}
\includegraphics[width=1.0\linewidth]{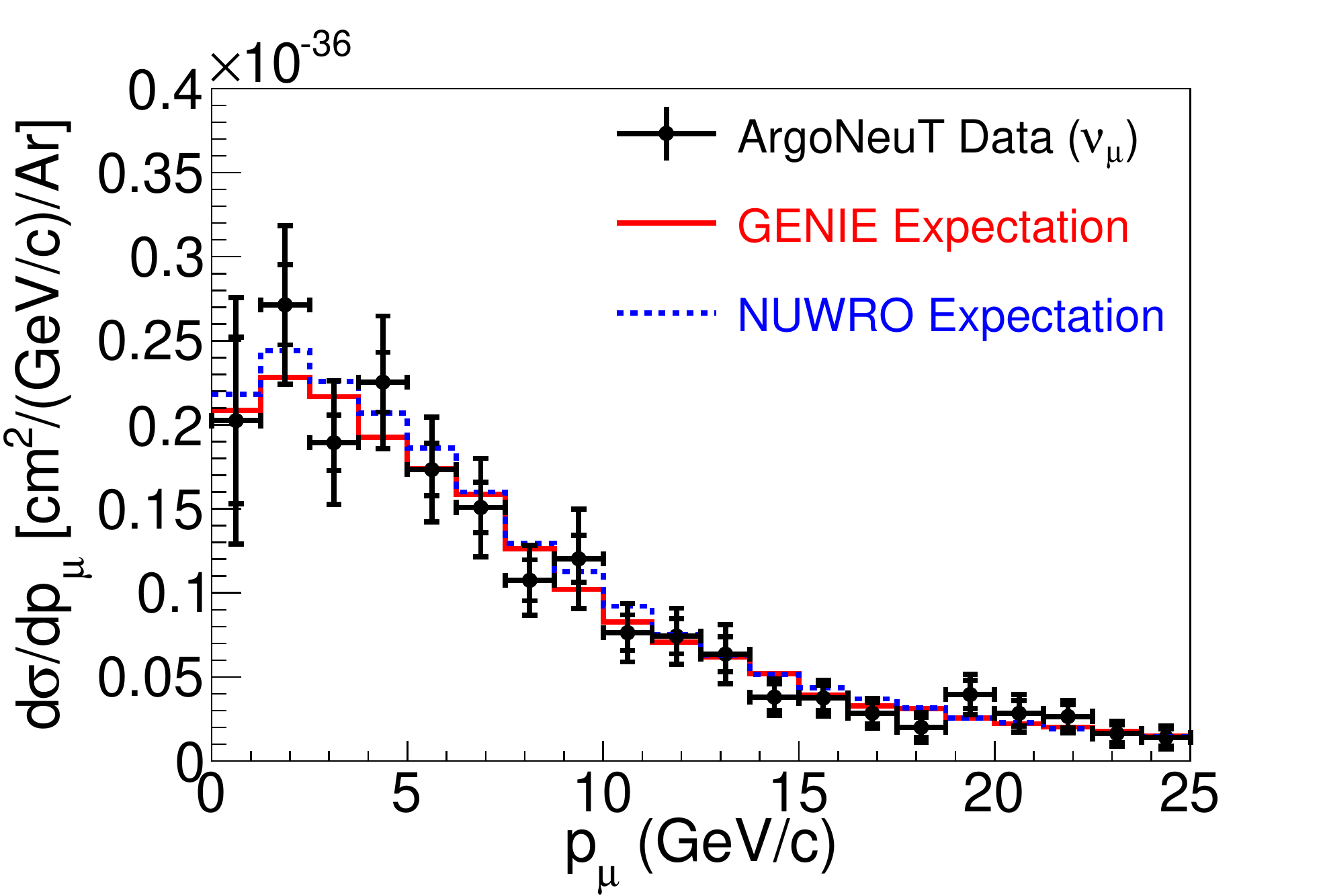}
\caption{The measured differential cross sections in muon (top) angle and (bottom) momentum for CC inclusive $\nu_{\mu}$ interactions in argon, per argon nucleus, in comparison with {\sc genie}~\cite{genie} and {\sc nuwro}~\cite{nuwro} expectations. Both statistical and total errors are shown on the data points.}
\vspace{-.5cm}
\label{fig:ardiff}
\end{figure}

\begin{figure}[htp]
\includegraphics[width=1.0\linewidth]{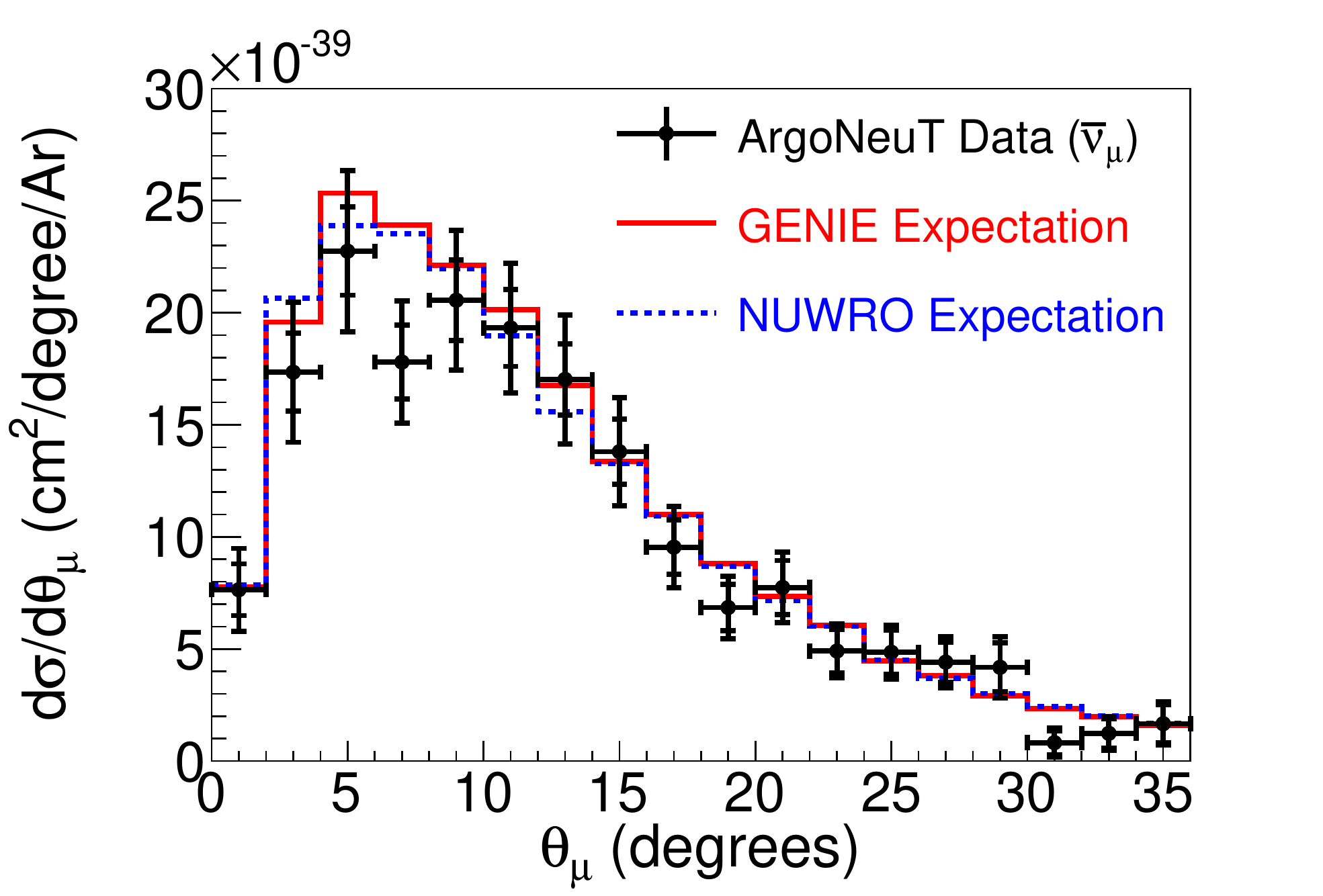}
\includegraphics[width=1.0\linewidth]{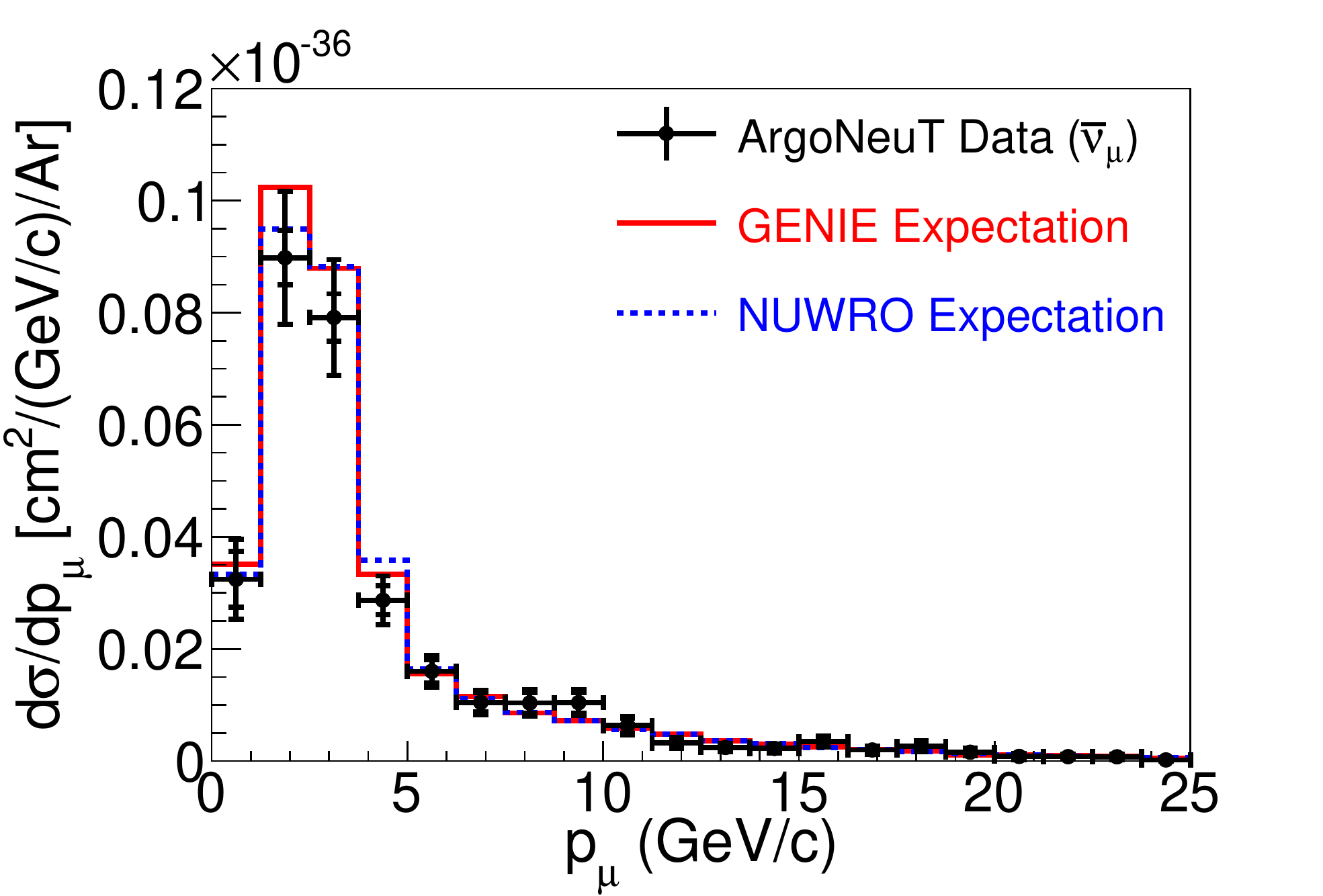}
\vspace{-.5cm}
\caption{The measured differential cross sections in muon (top) angle and (bottom) momentum for CC inclusive $\overline{\nu}_{\mu}$ interactions in argon, per argon nucleus, in comparison with {\sc genie}~\cite{genie} and {\sc nuwro}~\cite{nuwro} expectations. Both statistical and total errors are shown on the data points.}
\label{fig:ardiff2}
\end{figure}

The flux-integrated differential cross sections in $\theta_\mu$ and $p_\mu$ from $\nu_\mu$ ($\overline{\nu}_\mu$) CC events on an argon target are shown in Figs.~\ref{fig:ardiff} (\ref{fig:ardiff2}), and are tabulated in Tables.~\ref{tab:angle} and~\ref{tab:pmu}. 

\begin{table}[htp]
\centering
\caption{The measured differential cross sections in muon angle for CC $\nu_{\mu}$ and $\overline{\nu}_{\mu}$ interactions in argon, per argon nucleus. Both statistical and total errors are shown.}
\begin{tabular}{ccc}\hline \hline
$\theta_{\mu}$ & $d\sigma/d\theta_{\mu} (\nu_{\mu})$ & $d\sigma/d\theta_{\mu} (\overline{\nu}_{\mu})$  \\
(degrees) &($10^{-38}\mathrm{cm}^2/\mathrm{degree}$/Ar) & ($10^{-38}\mathrm{cm}^2/\mathrm{degree}$/Ar) \\  \hline
0-2 & $7.31 \pm 0.88 \pm 2.87$ & $0.76 \pm 0.11 \pm 0.18$ \\ 
2-4 & $16.7 \pm 1.3 \pm 3.5$ & $1.7 \pm 0.2 \pm 0.3$ \\ 
4-6 & $15.7 \pm 1.2 \pm 2.7$ & $2.3 \pm 0.2 \pm 0.4$ \\ 
6-8 & $16.0 \pm 1.2 \pm 2.9$ & $1.8 \pm 0.2 \pm 0.3$ \\ 
8-10 & $11.2 \pm 1.0 \pm 2.5$ & $2.1 \pm 0.2 \pm 0.3$ \\ 
10-12 & $8.86 \pm 0.90 \pm 1.78$ & $1.93 \pm 0.17 \pm 0.29$ \\ 
12-14 & $8.32 \pm 0.87 \pm 2.07$ & $1.70 \pm 0.16 \pm 0.29$ \\ 
14-16 & $6.74 \pm 0.83 \pm 1.41$ & $1.38 \pm 0.14 \pm 0.24$ \\ 
16-18 & $5.32 \pm 0.74 \pm 1.36$ & $0.95 \pm 0.12 \pm 0.18$ \\ 
18-20 & $4.49 \pm 0.66 \pm 1.12$ & $0.68 \pm 0.10 \pm 0.14$ \\ 
20-22 & $4.29 \pm 0.70 \pm 1.06$ & $0.77 \pm 0.12 \pm 0.16$ \\ 
22-24 & $3.96 \pm 0.74 \pm 1.06$ & $0.49 \pm 0.10 \pm 0.12$ \\ 
24-26 & $2.67 \pm 0.63 \pm 0.85$ & $0.48 \pm 0.10 \pm 0.12$ \\ 
26-28 & $1.23 \pm 0.43 \pm 0.54$ & $0.44 \pm 0.09 \pm 0.11$ \\ 
28-30 & $2.35 \pm 0.65 \pm 0.88$ & $0.42 \pm 0.11 \pm 0.13$ \\ 
30-32 & $1.04 \pm 0.44 \pm 0.56$ & $0.08 \pm 0.05 \pm 0.06$ \\ 
32-34 & $0.61 \pm 0.48 \pm 0.61$ & $0.12 \pm 0.06 \pm 0.07$ \\ 
34-36 & $1.38 \pm 0.64 \pm 0.77$ & $0.17 \pm 0.08 \pm 0.10$ \\ 
\hline\hline
\end{tabular} 
\label{tab:angle}
\end{table}

\begin{table}[htp]
\centering
\caption{The measured differential cross sections in muon momentum for CC $\nu_{\mu}$ and $\overline{\nu}_{\mu}$ interactions in argon, per argon nucleus. Both statistical and total errors are shown.}
\begin{tabular}{ccc}\hline\hline 
$p_{\mu}$ & $d\sigma/dp_{\mu} (\nu_{\mu}$) & $d\sigma/dp_{\mu} (\overline{\nu}_{\mu}$) \\
(GeV/c) & [$10^{-38}\mathrm{cm}^2/(\mathrm{GeV/c})$/Ar] & [$10^{-38}\mathrm{cm}^2/(\mathrm{GeV/c})$/Ar] \\  \hline
0-1.25 & $20.3 \pm 4.9 \pm 7.3$ & $3.2 \pm 0.5 \pm 0.7$ \\ 
1.25-2.5 & $27.1 \pm 2.3 \pm 4.7$ & $9.0 \pm 0.5 \pm 1.2$ \\ 
2.5-3.75 & $18.9 \pm 1.65 \pm 3.7$ & $7.9 \pm 0.4 \pm 1.0$ \\ 
3.75-5 & $22.5 \pm 1.8 \pm 3.9$ & $2.9 \pm 0.3 \pm 0.4$ \\ 
5-6.25 & $17.3 \pm 1.5 \pm 3.1$ & $1.6 \pm 0.2 \pm 0.3$ \\ 
6.25-7.5 & $15.1 \pm 1.5 \pm 2.9$ & $1.0 \pm 0.2 \pm 0.2$ \\ 
7.5-8.75 & $10.8 \pm 1.2 \pm 2.1$ & $1.0 \pm 0.2 \pm 0.2$ \\ 
8.75-10 & $12.0 \pm 1.4 \pm 3.0$ & $1.0 \pm 0.2 \pm 0.2$ \\ 
10-11.2 & $7.64 \pm 1.06 \pm 1.74$ & $0.63 \pm 0.14 \pm 0.16$ \\ 
11.2-12.5 & $7.43 \pm 1.05 \pm 1.66$ & $0.32 \pm 0.10 \pm 0.11$ \\ 
12.5-13.8 & $6.36 \pm 1.04 \pm 1.75$ & $0.24 \pm 0.07 \pm 0.10$ \\ 
13.8-15 & $3.80 \pm 0.78 \pm 1.10$ & $0.22 \pm 0.08 \pm 0.09$ \\ 
15-16.2 & $3.75 \pm 0.74 \pm 1.06$ & $0.34 \pm 0.10 \pm 0.11$ \\ 
16.2-17.5 & $2.84 \pm 0.64 \pm 0.89$ & $0.20 \pm 0.07 \pm 0.08$ \\ 
17.5-18.8 & $2.01 \pm 0.59 \pm 0.88$ & $0.27 \pm 0.10 \pm 0.12$ \\ 
18.8-20 & $3.96 \pm 0.83 \pm 1.19$ & $0.16 \pm 0.07 \pm 0.08$ \\ 
20-21.2 & $2.83 \pm 0.75 \pm 1.11$ & $0.08 \pm 0.05 \pm 0.06$ \\ 
21.2-22.5 & $2.65 \pm 0.69 \pm 0.96$ & $0.08 \pm 0.04 \pm 0.05$ \\ 
22.5-23.8 & $1.63 \pm 0.53 \pm 0.73$ & $0.08 \pm 0.05 \pm 0.06$ \\ 
23.8-25 & $1.40 \pm 0.50 \pm 0.67$ & $0.02 \pm 0.02 \pm 0.02$ \\ 
\hline\hline
\end{tabular} 
\label{tab:pmu}
\end{table}

Uncertainties associated with measurement resolution are evaluated by recalculating the differential cross sections under two different efficiency assumptions. The default efficiency takes into account smearing effect from detector resolution, and the other calculation assumes there is no smearing. The difference between these is taken as a systematic error on the measured differential cross sections.
The {\sc genie} predictions agree with data for neutrinos but overestimate data slightly for antineutrinos. {\sc nuwro}~\cite{nuwro} expectations are consistent with those from {\sc genie} in most bins.


The total cross section systematic error contributions are dominated by the uncertainty in the energy-integrated flux. The flux used in this analysis~\cite{minosflux} is based on a simulation of the NuMI beamline with the {\sc fluka}~\cite{Battistoni:2007zzb} hadron production tuned with MINOS near detector data~\cite{Adamson:2011fa} and NA49 hadron production measurements~\cite{Alt:2006fr}. We assign a 
flat 11\% flux error which accounts for the uncertainties in the hadron production and beamline modeling ({\it e.g.}~horn focusing) and is consistent with the error assignment recently chosen by MINERvA ~\cite{minervanu,minervaanu} at low energies. The fluxes with uncertainties used in this measurement are reported in Table~\ref{fluxtable}. All considered sources of systematic uncertainty on the total cross section and their contributions are shown in Table~\ref{systable}.

\begin{table}[htp]
\begin{center}
\caption{The neutrino and antineutrino fluxes corresponding to the differential cross section measurements. The flux unit is $\nu_{\mu}$/GeV/m$^2$/10$^9$ POT.}
\begin{tabular}{ccc}\hline \hline
E$_\nu$ (GeV) & $\nu_{\mu}$ Flux& $\overline{\nu}_{\mu}$ Flux \\
\hline
0-1&$2.4\times10^3$&$8.0\times10^3$\\
1-2&$3.3\times10^3$&$4.2\times10^4$\\
2-3&$3.7\times10^3$&$6.9\times10^4$\\
3-4&$3.9\times10^3$&$6.9\times10^4$\\
4-5&$3.9\times10^3$&$2.8\times10^4$\\
5-6&$4.1\times10^3$&$8.9\times10^3$\\
6-7&$4.3\times10^3$&$4.7\times10^3$\\
7-8&$4.0\times10^3$&$3.3\times10^3$\\
8-9&$3.6\times10^3$&$2.4\times10^3$\\
9-10&$3.1\times10^3$&$1.9\times10^3$\\
10-11&$2.7\times10^3$&$1.6\times10^3$\\
11-12&$2.2\times10^3$&$1.2\times10^3$\\
12-13&$1.8\times10^3$&$1.1\times10^3$\\
13-14&$1.5\times10^3$&$8.3\times10^2$\\
14-15&$1.3\times10^3$&$6.8\times10^2$\\
15-16&$1.1\times10^3$&$5.6\times10^2$\\
16-17&$9.4\times10^2$&$4.8\times10^2$\\
17-18&$7.7\times10^2$&$3.7\times10^2$\\
18-19&$6.9\times10^2$&$3.1\times10^2$\\
19-20&$5.7\times10^2$&$2.6\times10^2$\\
20-25&$3.8\times10^2$&$1.5\times10^2$\\
25-30&$2.0\times10^2$&$61$\\
30-35&$1.1\times10^2$&$29$\\
35-40&$69$&$14$\\
40-45&$50$&$7.5$\\
45-50&$42$&$4.4$\\
\hline\hline
\end{tabular}
\label{fluxtable} 
\end{center}
\end{table}

\begin{table*}[htp]
\begin{center}
\caption{The systematic error contributions to the neutrino and antineutrino total cross sections.}
\begin{tabular}{ccc}\hline \hline
Systematic Uncertainty & $\nu_{\mu}$ contribution (\%) & $\overline{\nu}_{\mu}$ contribution (\%) \\ \hline
Number of Ar Targets & $2.0$&  $2.0$  \\
POT & $1.0$& $1.0$  \\
Fiducial Volume & $2.0$ & $1.6$ \\
Flat Flux Uncertainty & $11.0$ & $11.0$  \\ 
\hline\hline
\end{tabular}
\label{systable} 
\end{center}
\end{table*}

The total integrated cross section per nucleon in this analysis is calculated for both neutrinos and antineutrinos. 
The measured total $\nu_\mu$ ($\overline{\nu}_\mu$) CC cross section is $\sigma/E_\nu = 0.66 \pm 0.03 \pm 0.08$   ($0.28 \pm 0.01 \pm 0.03$)
$\times10^{-38}$ cm$^2$/GeV
per isoscalar nucleon at $\langle E_\nu \rangle=9.6 (3.6) \pm 6.5 (1.5)$~GeV, where the first error is statistical and the second is systematic, and the $\pm 6.5 (1.5)$ GeV represents the range that contains 68\% of the flux. The argon-to-isoscalar correction has been applied in arriving at these results. The corrections are about -3\% for neutrinos and 3\% for antineutrinos evaluated using the {\sc genie} cross section model~\cite{genie}.  Smearing between bins in the differential cross sections does not affect the measurement of the total cross sections. 
These results and the previous ArgoNeuT result, which was measured at an average neutrino energy of 4.3 GeV~\cite{josh}, are consistent with total cross section measurements made by other experiments in a similar energy region~\cite{pdg}. 

ArgoNeuT reports the first measurement of $\overline{\nu}_\mu$ CC differential cross sections in argon based on 1.20$\times10^{20}$~POT collected in the NuMI antineutrino beam. The results are reported in terms of outgoing muon angle and momentum. Similar distributions for $\nu_{\mu}$'s in the beam are also provided. The results are useful for tuning neutrino event generators, reducing the systematics associated with a long baseline neutrino oscillation experiment's near-far comparison, and informing the theory of neutrino-nucleus interactions generally.

We gratefully acknowledge the cooperation of the MINOS collaboration in providing their data and simulation code for use in this analysis. We wish to acknowledge the support of Fermilab, the Department of Energy, and the National Science Foundation in ArgoNeuT's construction, operation, and data analysis.

\bibliographystyle{apsrev}   
\bibliography{ccincbib}

\end{document}